\documentclass[aps,prb,superscriptaddress,twocolumn,showpacs,natbib]{revtex4}

\usepackage{amsmath}
\usepackage{amssymb}
\usepackage{bm}
\usepackage{epsfig}
\usepackage{graphicx}
\usepackage{color}

\begin{document}

\title{Dynamics of optically injected currents in carbon nanotubes}

\author{ L. L. Bonilla}
\affiliation{Gregorio Mill\'an Institute for Fluid Dynamics, Nanoscience and Industrial Mathematics, Universidad Carlos III de Madrid,
Avenida de la Universidad 30, 28911 Legan\'es, Spain}
\author{M. Alvaro}
\affiliation{Gregorio Mill\'an Institute for Fluid Dynamics, Nanoscience and Industrial Mathematics, Universidad Carlos III de Madrid,
Avenida de la Universidad 30, 28911 Legan\'es, Spain}
\author{M. Carretero}
\affiliation{Gregorio Mill\'an Institute for Fluid Dynamics, Nanoscience and Industrial Mathematics, Universidad Carlos III de Madrid,
Avenida de la Universidad 30, 28911 Legan\'es, Spain}

\author{E. Ya. Sherman}
\affiliation{Department of Physical Chemistry, The University of the Basque Country, 48080 Bilbao, Spain}
\affiliation{IKERBASQUE Basque Foundation for Science, Bilbao, Spain}

\begin{abstract}
We consider theoretically the dynamics of electric currents optically injected in carbon nanotubes. 
Although the plasma oscillations
are not seen in these systems, the main effect on the carrier's motion
is due to strongly nonuniform space-charge Coulomb forces produced by time-dependent separation 
of injected electron and hole densities.
We calculate evolution of the dipole moment characterizing the time- and coordinate-dependent charge
density distributions and analyze different regimes of the dynamics. 
The developed time-dependent dipole moment leads to a dipole radiation 
in the THz frequency range for typical parameters of injected currents. 
\end{abstract}


\maketitle

\setcounter{equation}{0}

\section{Introduction}

Optical manipulation of carriers in bulk solids and artificial structures  
is an interesting problem for fundamental and applied physics.
By interference of single- and two-photon optical transitions induced by highly coherent 
laser beams, which can be controlled by the beams' phases, one can inject optically currents in semiconductors
\cite{Hache,Kerachian} and semiconductor nanostructures \cite{Smirl,Najmaie}. 
Recently, current injection using the same principle of interference
was reported for a different class
of systems such as graphene \cite{Norris} and
carbon nanotubes \cite{Newson}. The understanding of the following dynamics of the injected currents
can provide a valuable information both about the injection process and
interactions in the system. 

For two-dimensional semiconductor quantum wells \cite{Sherman} it was shown that the space-charge
effects due to nonuniform charge density play the crucial role in the electron motion while relatively heavy
holes can be taken at rest. If the space-charge effects dominate in the 
charge dynamics, the timescale of the evolution is given by the characteristic expected plasma frequency 
corresponding to the injected charge density and the laser spot size. However, the plasma oscillations should not 
be seen there since a highly nonuniform charge density is formed on the timescale of the order of the 
expected inverse plasma frequency.  

In this respect, carbon nanotubes are strongly different
from conventional semiconductors. {Semiconducting nanotubes, where carriers have finite effective masses,
were intensively investigated by optical techniques. Near the absorption threshold they demonstrate 
excitonic effects in the optical absorption spectra \cite{Korovyanko} and in the subsequent dynamics \cite{Ruzicka}.
Here electrons and holes give the same contribution to the optical properties. Another type of nanotubes
is metallic systems, where the {dispersion relation} 
of carriers is linear in the momentum, making the plasma frequency  
a poorly defined quantity. As a result, even {when the carrier momentum changes due to the relaxation and external forces,}
the velocity, and therefore the current, can remain constant. To change the carrier velocity 
{the momentum has to change sign.} In general, for these ``relativistic'' spectra, 
even relatively strong Coulomb forces do not lead to 
formation of excitons (for an exception, see Ref.[\onlinecite{Wang}]). 
This new type of dynamics, which will be of our interest here, can experimentally be 
seen in bunches of nanotubes containing metallic and semiconducting species. As we are interested in the
optical response in the relatively low frequency infrared domain, semiconducting nanotubes will not contribute {to}
 the properties of our interest. Moreover, metallic nanotubes can be separated from the semiconducting ones \cite{review},
to provide a system for experimental study of the effects considered here.}

A microscopic theory of current injection in {semiconducting} nanotubes has been developed \cite{Knorr} by
using the analysis of the transition matrix elements on the atomic scale.
However, the stage of {the subsequent} dynamics with a strongly nonuniform density has not yet been studied and understood.
Here we study this process. The time-dependent injected current is accompanied by
emission of radiation in the THz frequency domain. As we will show, the spectrum of this radiation provides information
about the dynamics and properties of the system.

\section{Model dynamics equations}

\begin{figure}[h!]
\begin{center}
\includegraphics[width=80mm]{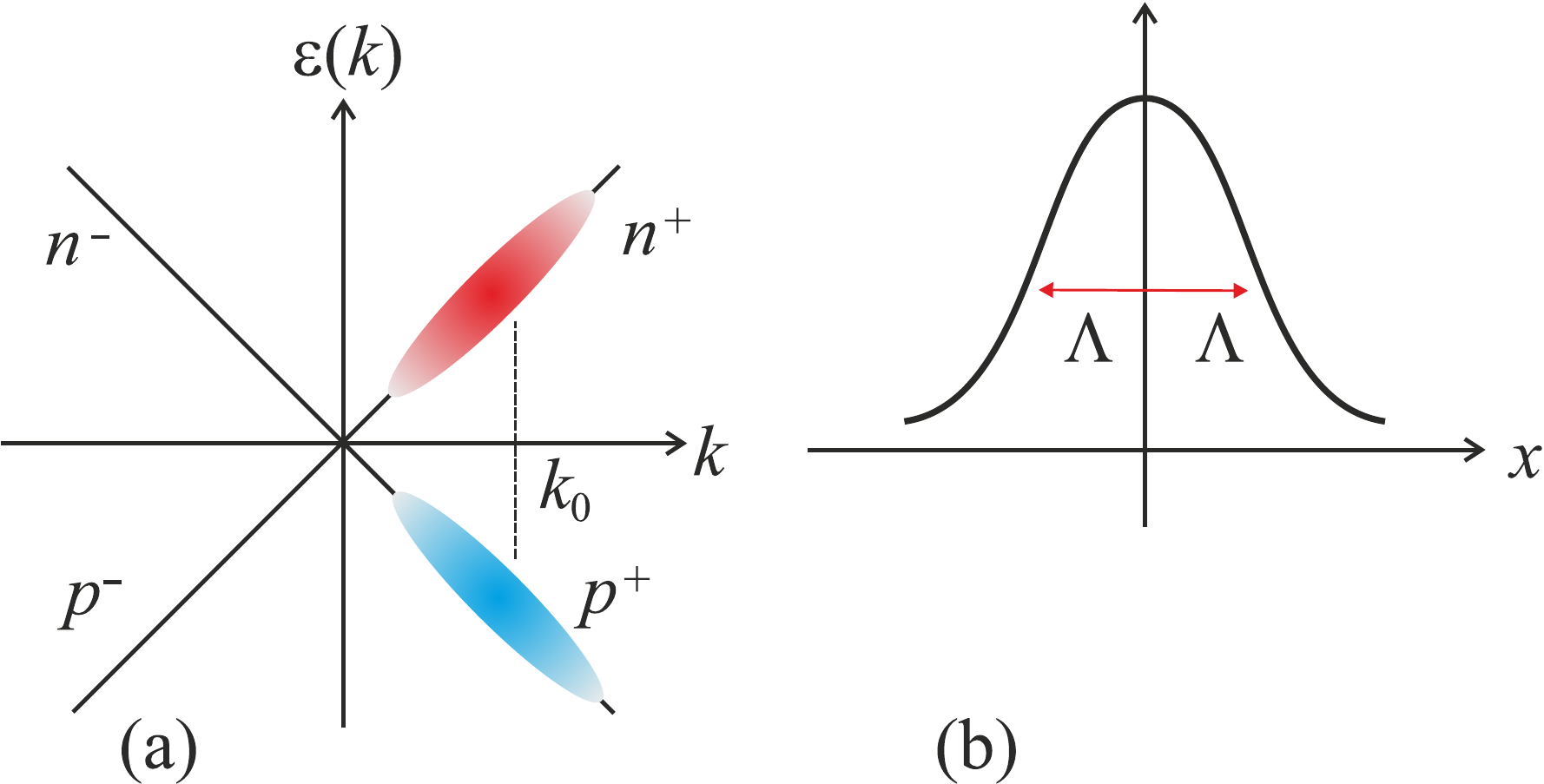}
\end{center}
\caption{(a) Dispersion relation and density distribution in the momentum space. The 
peak in the optical radiation intensity corresponds to transitions at frequency $2v_{0}k_{0}.$
(b) Electron/hole density induced by the laser spot vs $x$, where $2\Lambda$ is the total spot width.} \label{Figure1}
\end{figure}

We begin with model equations for a single wall carbon nanotube
characterized by dispersion relation (see Fig. \ref{Figure1}):
\begin{equation}
\varepsilon(k) = \hbar |k| v_{0}, \qquad \varepsilon(k) = -\hbar |k| v_{0},
\end{equation}
for electrons and holes, respectively, and velocities $v(k)=\pm\textrm{sign}(k)\, v_{0}$ where $v_{0} = 10^8$ cm/s.

Optical injection produces electron and hole
densities $n^{\pm}$, $p^{\pm}$ respectively in the coordinate $(x)$ and momentum $(k)-$ space wit corresponding 
velocities $v$:
\begin{eqnarray}
n^+ &=& n^+(x,k,t);\quad p^- = p^-(x,k,t); \quad v = v_{0};\\
n^-  &=& n^-(x,k,t);\quad p^+ = p^+(x,k,t);\quad v = -v_{0};
\end{eqnarray}
The local densities are defined as:
\begin{eqnarray}
\bar{n}^\pm = \bar{n}^\pm(x,t) = \int_{-\infty}^{+\infty} n^\pm(x,k,t) \,{d}k,\,\,\quad \bar{n}=\bar{n}^++\bar{n}^-;\\
\bar{p}^\pm = \bar{p}^\pm(x,t) = \int_{-\infty}^{+\infty} p^\pm(x,k,t) \,{d}k,\,\,\quad \bar{p}=\bar{p}^++\bar{p}^-.
\end{eqnarray}
In what follows we omit the explicit $(x,t,k)-$dependence for brevity.
The Boltzmann equations for the distribution functions have the form 
\begin{equation}
\label{eqdim1}
\frac{\partial n^{\pm}}{\partial t} + v(k) \frac{\partial n^{\pm}}{\partial x}
+\frac{e E}{\hbar} \frac{\partial n^{\pm}}{\partial k} = -\frac{n^\pm-n^\pm_{eq}}{\tau_n},
\end{equation}
\begin{equation}
\label{eqdim2}
\frac{\partial p^{\pm}}{\partial t} - v(k) \frac{\partial p^{\pm}}{\partial x} -\frac{e E}{\hbar} \frac{\partial
p^{\pm}}{\partial k} = -\frac{p^\pm-p^\pm_{eq}}{\tau_p},
\end{equation}
where $e<0$ is the electron charge and $E\equiv E(x,t)$ is the coordinate- and time-dependent electric field produced by
space-charge effects, that is by nonuniform charge density. The local Fermi-Dirac equilibrium
densities for electrons in Eq. (\ref{eqdim1}) and Eq. (\ref{eqdim2}) are defined as
\begin{equation}
n^\pm_{eq}=2\times\frac{1}{\exp\left[(\varepsilon(k)-\mu_{n}^{\pm})/T\right]+1},
\end{equation}
where the factor $2$ is due to spin degeneracy and $T$ is the temperature measured in units of energy.
In a similar way we define the equilibrium distributions of holes. The coordinate- and time-dependent 
chemical potentials $\mu_{n}^{\pm}$ and $\mu_{p}^{\pm}$
guarantee the balance of the densities in the form $\bar{n}^{\pm}_{eq}=\bar{n}^\pm$,
$\bar{p}^{\pm}_{eq}=\bar{p}^{\pm}$. For example, condition
\begin{equation}
\int_{0}^{\infty}n^{+}_{eq}{d}k=\bar{n}^{+},
\end{equation}
 yields
\begin{equation}
\mu_{n}^{+}= T\ln\left[\exp\left(\hbar v_0\bar{n}^{+}/2T\right)-1\right].
\end{equation}

The electric field in Eqs. (\ref{eqdim1}) and (\ref{eqdim2}) 
can be expressed in terms of the integral of the charge density
$\bar{p}(x-s,t)-\bar{n}(x-s,t)$ as:
\begin{equation}
E(x,t) = \int_{-\infty}^{\infty}
\left[\bar{p}(x-s,t) -\bar{n}(x-s,t)\right]{\cal K}(s)\,{d}s,
\label{Eformula}
\end{equation}
where ${\cal K}(s)$ is the Coulomb kernel for the nanotube.
The expression for ${\cal K}(s)$ is presented in the Appendix. As we will see, the
important and unusual feature of Eq. (\ref{Eformula}) is that in the limit of a small-radius nanotube the
field $E(x,t)$ is proportional to the local 
derivative of the density: $\partial\left(\bar{p}(x,t)-\bar{n}(x,t)\right)/\partial x.$

The initial distributions of electrons and holes are optically produced as:
\begin{equation}
n(x,k,0) = p(x,k,0) = \frac{N}{\pi \Lambda K}
\exp\!\!\left[-\frac{x^2}{\Lambda^2}-\frac{(k-k_0)^2}{K^2}\!\right],
\label{eqnp0}
\end{equation}
where $k_0$ is  the injection point in the momentum space,
$2\Lambda$ is the characteristic laser spot size, and $N$ is the total number of injected electron/holes.
We use {$k_0=K=200$ $\mu$m$^{-1}$ (see Fig. \ref{Figure1}(a)). This is reasonable since the wave vector is limited
by the requirement that the carrier energy should not exceed that of the optical phonon, 
for otherwise a fast momentum and energy relaxation occur. A typical
optical phonon energy is $\hbar\Omega_{\rm ph}=0.18$ eV, \cite{Piscanec} 
that gives the estimate $k_{0}<\Omega_{\rm ph}/v_{0}=275$ $\mu$m$^{-1}$.
Integrating Eq.(\ref{eqnp0}) over $k$, we obtain (see Fig. \ref{Figure1}(b)):
\begin{equation}
\bar{n}(x,0) = \bar{p}(x,0) = \frac{N}{\sqrt{\pi}\Lambda }
\exp\left(-x^2/\Lambda^2\right). \label{eqnp1}
\end{equation}
We define the one-dimensional (1D)
electron/hole density
\begin{equation}
N_{1D} = \frac{N}{2\Lambda},
\end{equation}
and Eq. (\ref{eqnp1}) becomes:
\begin{equation}
\bar{n}(x,0) = \bar{p}(x,0) = \frac{2N_{1D}}{\sqrt{\pi} }
\exp\left(-x^2/\Lambda^2\right).
\end{equation}

Integrating Eqs. (\ref{eqdim1}) and (\ref{eqdim2}) over $k$, we get the charge continuity equation:
\begin{eqnarray}
&&\frac{\partial}{\partial t} \left[ (\bar{p}^{+} + \bar{p}^{-}) - (\bar{n}^{+} + \bar{n}^{-}) \right] \\
&&\hspace{1cm}+v_{0} \frac{\partial}{\partial x} \left[(\bar{p}^{-}-\bar{p}^{+}) + (\bar{n}^{-}-\bar{n}^{+}) \right] = 0, \nonumber
\end{eqnarray}
{from which we} define the local current:
\begin{equation}
 I= I(x,t) = -e\, v_{0}\,\left[(\bar{p}^{-}-\bar{p}^{+})+(\bar{n}^{-}-\bar{n}^{+})\right]\,.
\end{equation}

\section{Numerical solutions}

Before solving numerically the model equations, we introduce parameters describing the
electron-electron interaction and the injection process. First we introduce a parameter
characterizing the strength of the Coulomb forces. For this purpose we use the following scaling
argument. The Coulomb force acting at a carrier, $F\sim Ne^{2}/\sqrt{\epsilon_{\perp}\epsilon_{\|}}\Lambda^{2}$ 
where $\epsilon_{\perp}$ and $\epsilon_{\|}$ is the nanotube transversal and longitudinal permittivity, respectively
(see Appendix for details).
{In the absence of the plasma frequency and on the relevant}
time scale $t_{\Lambda}\sim\Lambda/v_{F}$, this force produces a change in the 
momentum comparable to $\hbar K$ if $Ft_{\Lambda}\sim \hbar K$. This estimate yields the corresponding
critical number of injected carriers per nanotube $N_c\equiv\sqrt{\epsilon_{\perp}\epsilon_{\|}}\hbar v_{0}K\Lambda/{e^2}\approx 1600$ 
(for $K=200$ $\mu$m$^{-1}$, $\epsilon_{\perp}=10$ and $\epsilon_{\|}=30,$ in agreement 
with the experiment [\onlinecite{Lu}]), and 
the interaction effects are described by a dimensionless parameter $N/N_{c}.$
If ${N}/{N_c}\ll 1$, then $\bar{n}^{\pm}$ and $\bar{p}^{\pm}$ are conserved
and the initial density distributions move and separate being only weakly deformed. Otherwise, the effect
of Coulomb forces is strong. In this case, if $E<0$ ($E>0$)  then $\bar{n}^{+}$ and $\bar{p}^{-}$
increase (decrease), and $\bar{p}^{+}$ and $\bar{n}^{-}$ decrease (increase). Since initially
carriers are injected with positive momentum $k_0>0$ most of the electrons/holes have $k>0$.
Thus $n(x,0,t)$ increases and $p(x,0,t)$ decreases when $E>0$, and therefore the effect of
the electric field is larger on the electrons than on the holes. Conversely, when $E<0$ the
effect of the electric field is larger on the holes than on the electrons. However, when the
number of electrons/holes having $k<0$ is larger, the sign of the electric field has the
opposite effect for those electrons/holes.

An important limit on the number of injected particles is  posed by the Pauli blocking condition, where the injection stops since all available
electron/hole states became occupied by the previously excited carriers. {The condition that the
transition does not experience Pauli blocking limits the number of injected particles to the
available phase volume $2\Lambda\Delta K$, where $2$ is the spin factor. Therefore, the maximum
ratio $N/N_{c}$ should be considerably less than 
$(\Delta K/K)\alpha(c/v_{F})(2/\sqrt{\epsilon_{\perp}\epsilon_{\|}})$, where $\alpha\equiv e^{2}/\hbar c=1/137$ is the fine
structure constant. Provided $\Delta K\sim K=275$ $\mu$m$^{-1}$, we obtain that $N$ should be less than 400
restricting $N/N_c$ to values considerably less than 0.3.

We consider current injection by tightly focused beams with $\Lambda= 1 \,\mu$m, and nanotube of the radius $a=1.25$ and 
solve numerically the equations
(\ref{eqdim1}) to (\ref{Eformula}) in the following two cases (see Fig. \ref{Figure4}--Fig. \ref{Figure8}): 
(a) $N_{1D}\approx 1.25\times 10^5$ cm$^{-1}$, injected $N= 25$,
$N/N_c\approx 0.016$, and $I_{\rm max}\approx 4$ $\mu$A and (b) $N_{1D}\approx 1.25\times 10^6$
cm$^{-1}$, $N= 250$, $N/N_c\approx 0.16$, and $I_{\rm max}\approx 40\,\mu$A. In both cases we
assume carriers scattering times $\tau_{n}=\tau_{p}=2$ ps at 300 K, as suggested by the 
estimates \cite{Lazzeri,Sundqvist}.

\begin{figure}[h!]
\begin{center}
  \includegraphics[width=80mm]{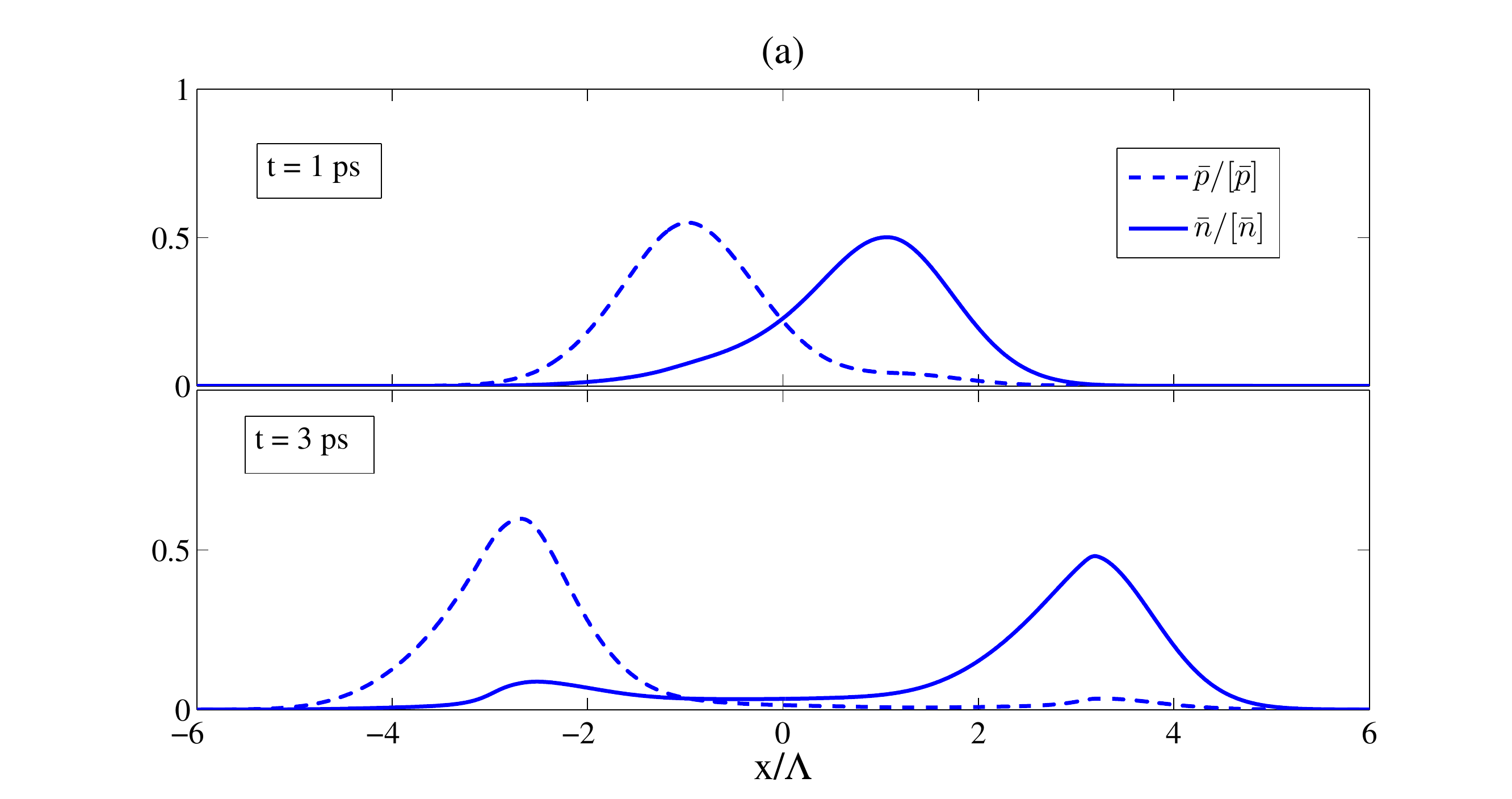} \\
  \includegraphics[width=80mm]{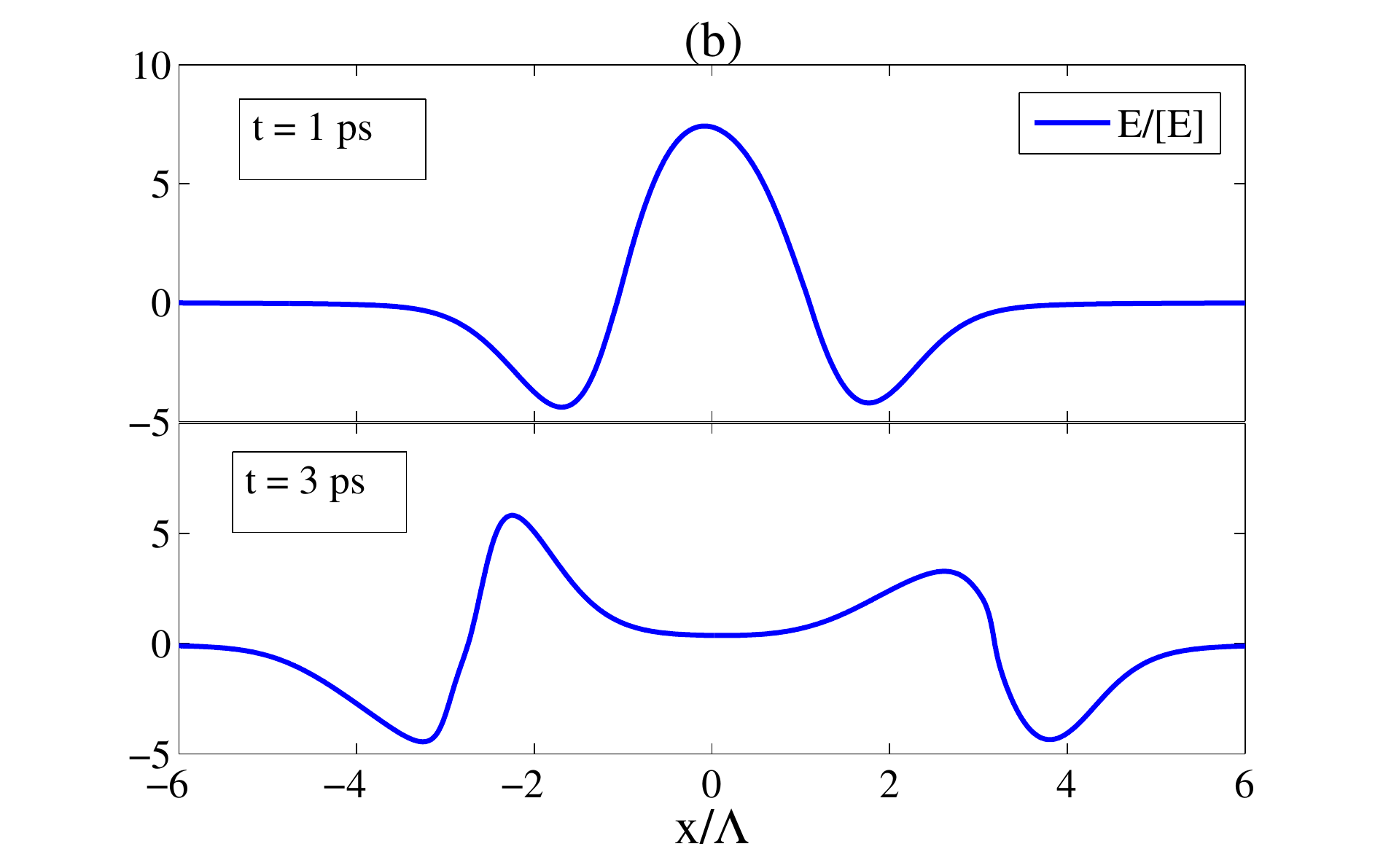}
\end{center}
\caption{Profiles of (a) density of electrons/holes, and (b) electric field at different snapshots
for case a):  $\Lambda=1 \,\mu$m, $N_{1D}\approx 12.5$ $\mu$m$^{-1}$, $\tau_p=\tau_n=2$ ps, $N=25$
electrons/holes. {Here $[\bar{p}]=[\bar{n}]=N/\Lambda= 2.5\times 10^{5}$ cm$^{-1}$, and the unit of electric field 
$[E]\equiv|e|N /\epsilon_{\perp} \Lambda^2= 36$ V/cm. }} 
\label{Figure4}
\end{figure}

In case (a), there are few carriers and electrons and holes go their separate ways without much
interaction, as shown in Fig. \ref{Figure4}. The effect of the
non-equilibrium electric field is much greater in case (b), when there are ten
times more carriers. Figure \ref{Figure5} shows that the interaction between carriers
builds up a peak in the hole density. The extrema of the electric field are reached at
the inflection points of the charge density $\bar{p}(x,t)-\bar{n}(x,t)$ in agreement
with the approximate formulas presented in the Appendix. The displacement and separation 
of electron and hole peaks is of the order of the initial width $\Lambda$, that is much larger 
than can observed in semiconductor quantum wells \cite{Smirl,Sherman}. The reduction 
of separation by space-charge effects due to the finite mass of the carriers, similar to that in 
the quantum wells, can be expected in semiconductor nanotubes as well.
As one can see in the Figures for carrier densities and electric fields,
the scattering sharpens the peaks in
the carrier densities and the electric field and depresses the smooth regions 
thereof but does not change this qualitative picture. 
The reason is that the scattering tries to keep the carrier 
densities close their local equilibrium values (Fermi functions) which have 
large gradients near $k=0$. This enhances the effect of the electric field on the densities and sharpens their peaks.

\begin{figure}[h!]
\begin{center}
\includegraphics[width=80mm]{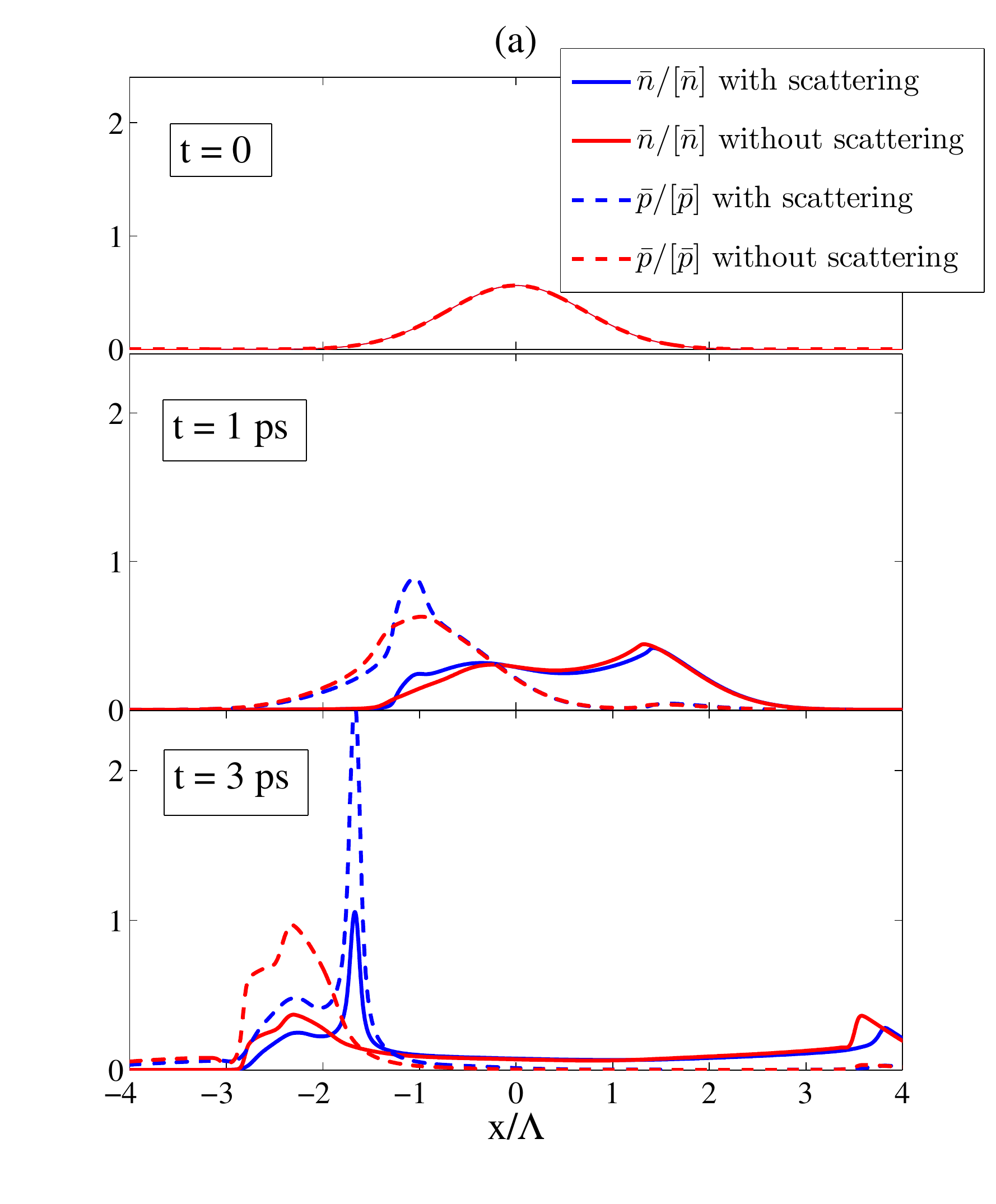}\vspace{-5mm} \\
\includegraphics[width=80mm]{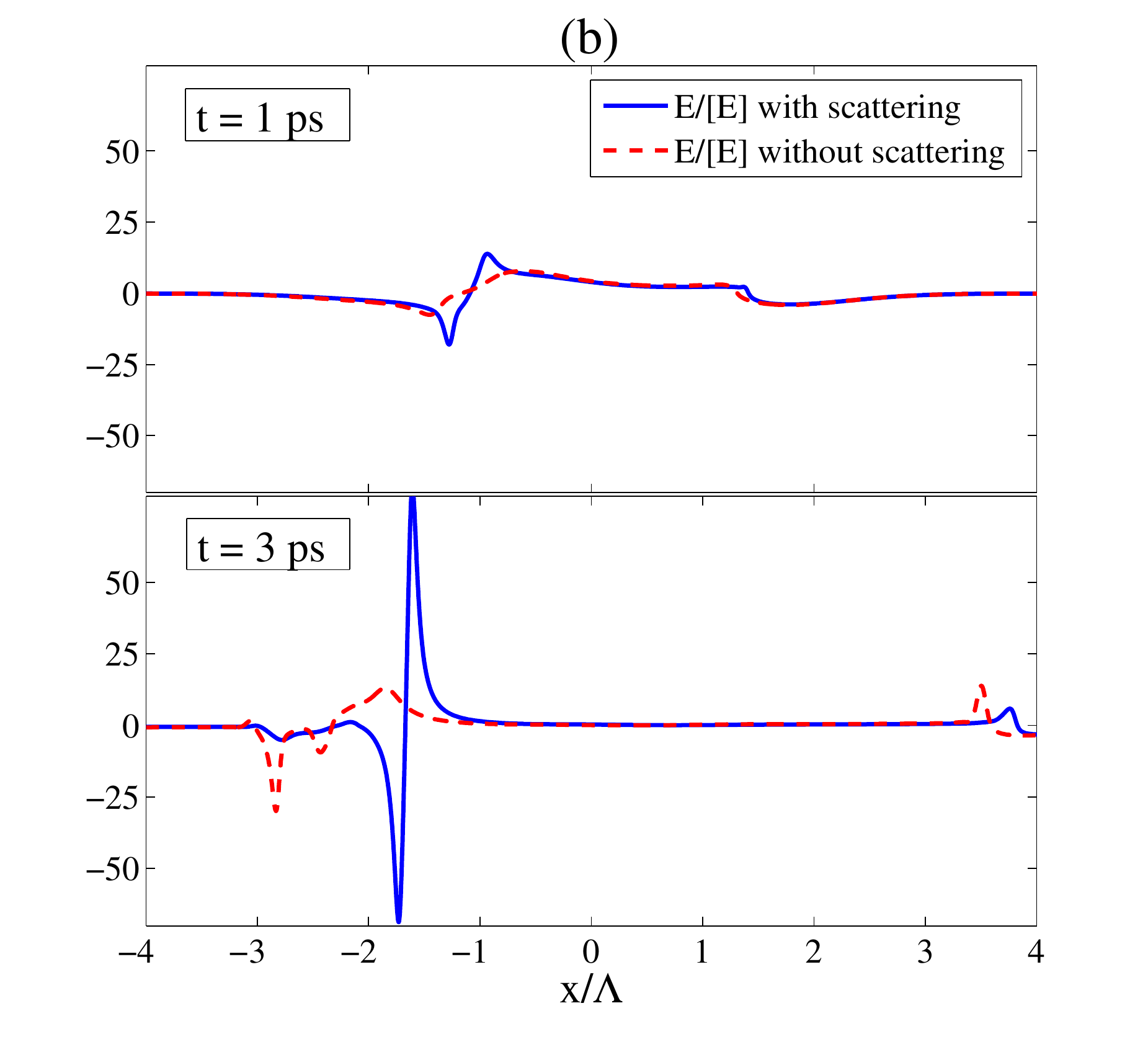}\vspace{-5mm}
\end{center}
\caption{ (a) Profiles of the electron/hole densities, and (b) electric field profile, at different snapshots, for $N=250$ electrons/holes.
Here $[\bar{p}]=[\bar{n}]=N/\Lambda= 2.5\times 10^{6}$ cm$^{-1}$, and the unit of electric field $[E]\equiv|e|N /\epsilon_{\perp} \Lambda^2= 360$ V/cm. 
} 
\label{Figure5}
\end{figure}

Figure \ref{Figure5}(b) shows the development of electric fields and, thus,
details the way the carriers peaks are built. Since we inject current with
positive momentum, the larger peaks of electron and hole densities correspond to $k>0$.
However, the hole population with $k<0$ splits in two parts and one part moves together
with the hole population with $k>0$ which helps building up the hole population at the
peak that moves to the left. Meanwhile, the electron population with $k>0$ also splits in
two parts and the one that moves to the left helps reinforcing the electron population
with $k<0$. The location of the electron peak that moves to the left is quite close 
to that of the left-moving hole peak. Then electrons and holes interact so that their left-moving peaks
slow down almost to a halt at the same location.

\begin{figure}[h!]
\begin{center}
 \includegraphics[width=80mm]{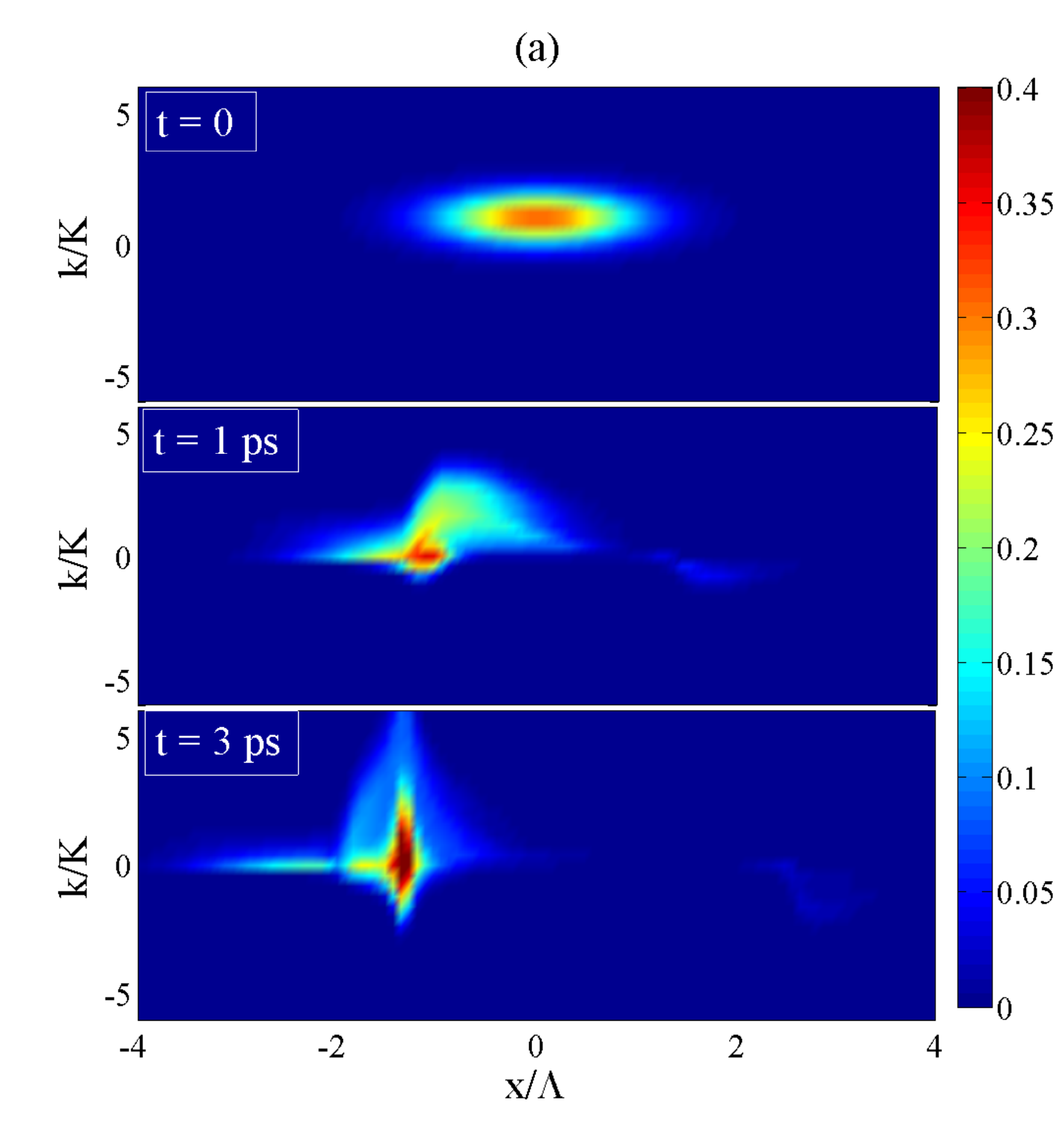}\\
 \includegraphics[width=80mm]{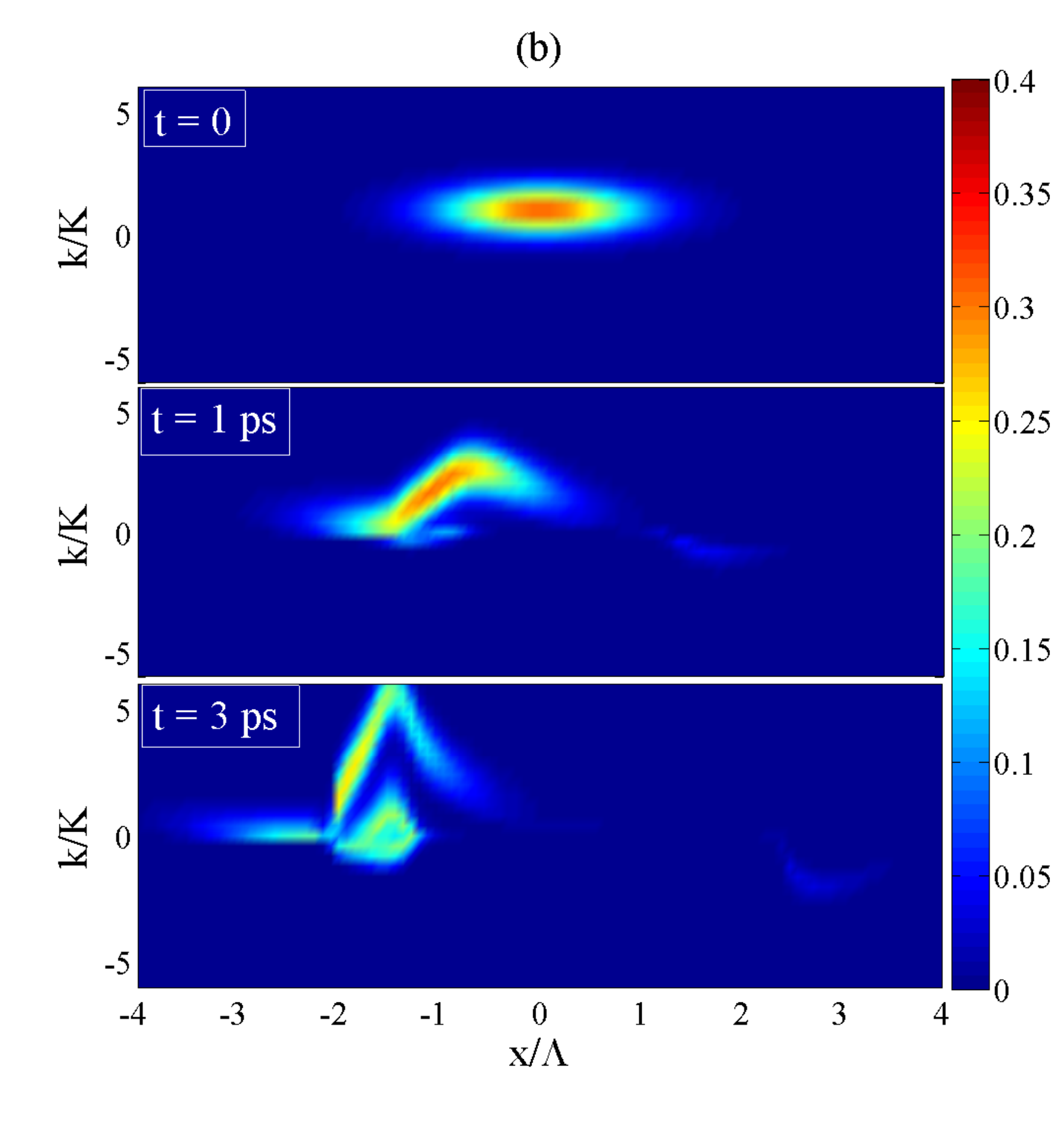}
\end{center}
\caption{Hole density $(p^+(x,k)+p^-(x,k))[p]$ at different snapshots for $N=250$
electrons/holes: (a) with scattering,  (b) without scattering. Here $[p]=N/\Lambda K=
1.25$.} \label{Figure7}
\end{figure}

This picture is confirmed in Fig. \ref{Figure7} that shows snapshots of the overall hole density $p^+(x,k,t)+p^-(x,k,t)$
for $N=250$ electron/holes.
Note that the Coulomb forces stop the motion of carriers
to the left and build up hole and electron peaks at $x\approx 2.3\Lambda$.
Similarly, increasing the number of carriers narrows the peaks of their spatial 
density distributions, as shown in Fig. \ref{Figure8}.

\begin{figure}[t]
\begin{center}
 \includegraphics[width=80mm]{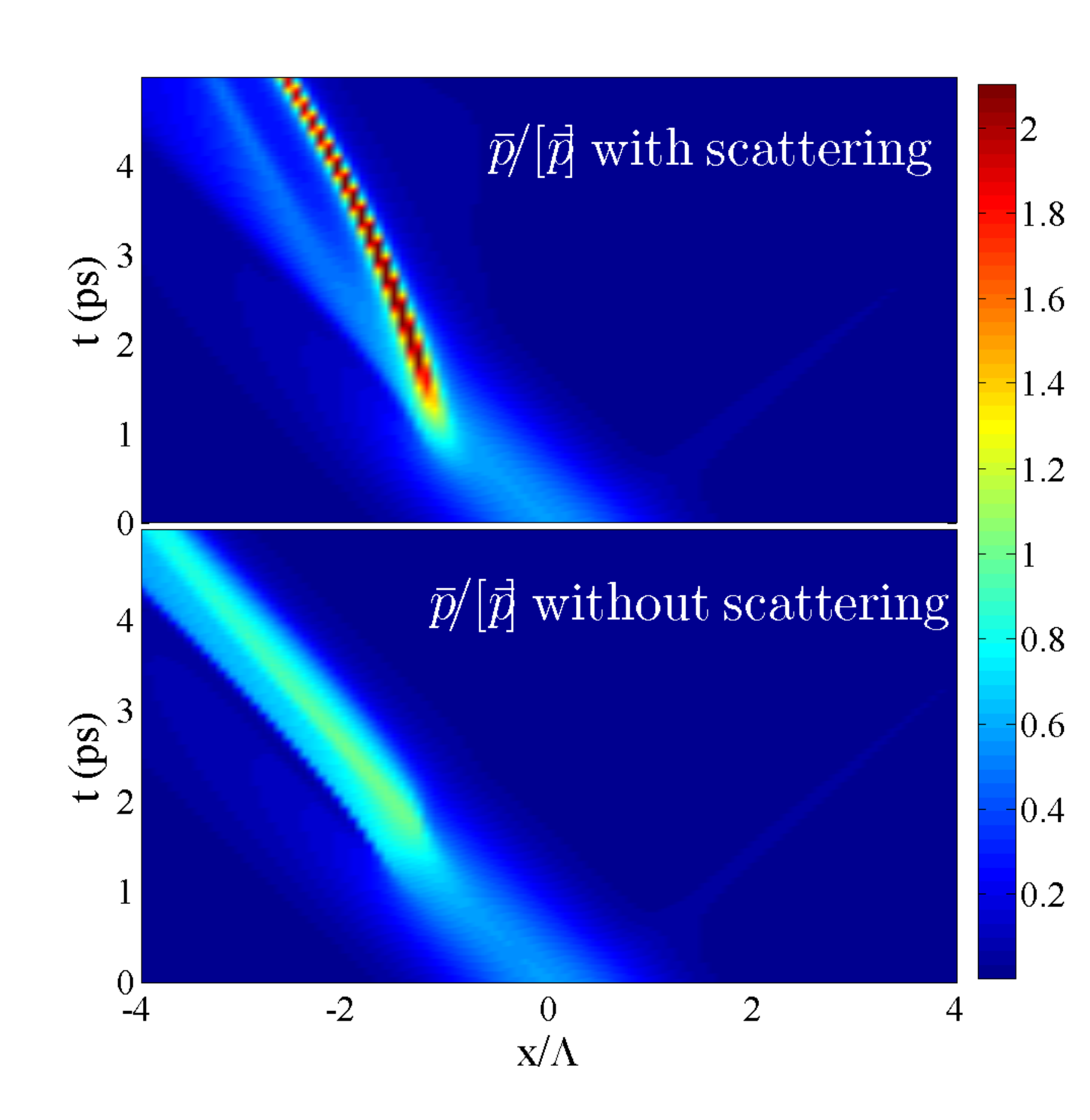}
\end{center}
\caption{$\bar{p}(x,t)/[\bar{p}]$ for $N=250$ electrons/holes. Here $[\bar{p}]=N/\Lambda=
2.5\times 10^{6}$ cm$^{-1}$.} \label{Figure8}
\end{figure}

\section{Dipole radiation: intensity and spectrum}

In order to make relation of our results to possible experimental observations of the space-charge effects in the current evolution, 
here we study the radiation of a single nanotube after the current injection to see how
the system parameters can be found from the radiation intensity and the spectrum.
For this purpose we introduce the dipole moment as:
\begin{equation}
D(t) = -e\int_{-\infty}^{+\infty} (\bar{p}(x,t)-\bar{n}(x,t))x \,{d}x,
\end{equation}
where the corresponding radiation intensity is proportional to $\left({d^2 D}/{d t^2}\right)^2$.
The maximum value of $D$ can be estimated as $N|e|\Lambda.$ Taking into account that the 
timescale of the process is $\Lambda/v_{0}$, we obtain the typical value of ${d^2 D}/{d t^2}$ 
of the order of $|e|v_{0}^{2}\times N/\Lambda.$ Taking into account
the Pauli blocking limitations, $N/\Lambda\le K_{0}$, we obtain the fundamental limit for the derivative
${d^2 D}/{d t^2}\le ev_{0}\epsilon(k_{0})/\hbar.$
The corresponding spectral density for current injected at $t=0$ is given by
\begin{equation}
I(\omega)\sim\left|\int_{0}^{+\infty}\frac{d^2 D}{d t^2}e^{i\omega t}dt\right|^{2}.
\end{equation}
Figure \ref{Figure9} shows ${d^2 D}/{d t^2}$ for $N=250$ with and without scattering effects.
The system parameters demonstrate themselves in the spectrum of the radiation. 
Increasing the number of carriers produces a
sharp peak which is somewhat augmented and sharpened by scattering. 
Although the scattering sharpens the distributions, it only weakly 
modifies the integral parameter such as the dipole moment, as can be seen in Fig. \ref{Figure9}.
The Fourier transform of ${d^2 D}/{d t^2}$ provides the spectrum of the radiation peaked 
at the frequency of $200$ GHz for a time window of 5 ps.

\section{Conclusions}

We have studied the time evolution of charge density 
after optical injection of a charge current in metallic carbon nanotubes with ``relativistic'' spectrum and identified 
different regimes of dynamics. The main impact on the carrier density evolution is produced by the space-charge
effects. However, due to the zero effective mass of the carriers, these Coulomb
forces cannot prevent the calculated large separation of the electron and 
hole densities. This is in contrast to the relatively small separation expected 
in semiconductor structures, where electrons and holes have finite masses. 
Although the scattering of carriers by impurities and phonons 
considerably sharpens the density distribution, it does not influence strongly its
integral characteristics such as the dipole moment resulting from the electron-hole separation, thereby 
rendering difficult the experimental verification of this effect. The time evolution
of the dipole moment leads to a dipole radiation, which can be measured 
experimentally and provide information about the dynamics of the carriers. The spectral width of the radiation is mainly determined by the 
ratio of the ``relativistic'' velocity to the spatial width of the initial density distribution, while the intensity depends on the 
injected carrier density. Our results show that, as the intensity of the exciting radiation increases, there is a fundamental limit for the radiation intensity that is determined solely by the radiation frequency and related to Pauli blocking
in the injection process. Realistic numerical parameters correspond to the radiation spectrum peaked at a fraction of a THz,
in the range of experimental observation of Ref.[\onlinecite{Newson}].

\section{Acknowledgement} This work was supported by the University of Basque Country UPV/EHU under
program UFI 11/55, Spanish MICINN (projects FIS2011-28838-C02-01 and FIS2012-36673-C03-01), and ``Grupos Consolidados UPV/EHU del Gobierno Vasco'' (IT-472-10).

\begin{figure}[t]
\begin{center}
  \includegraphics[width=80mm]{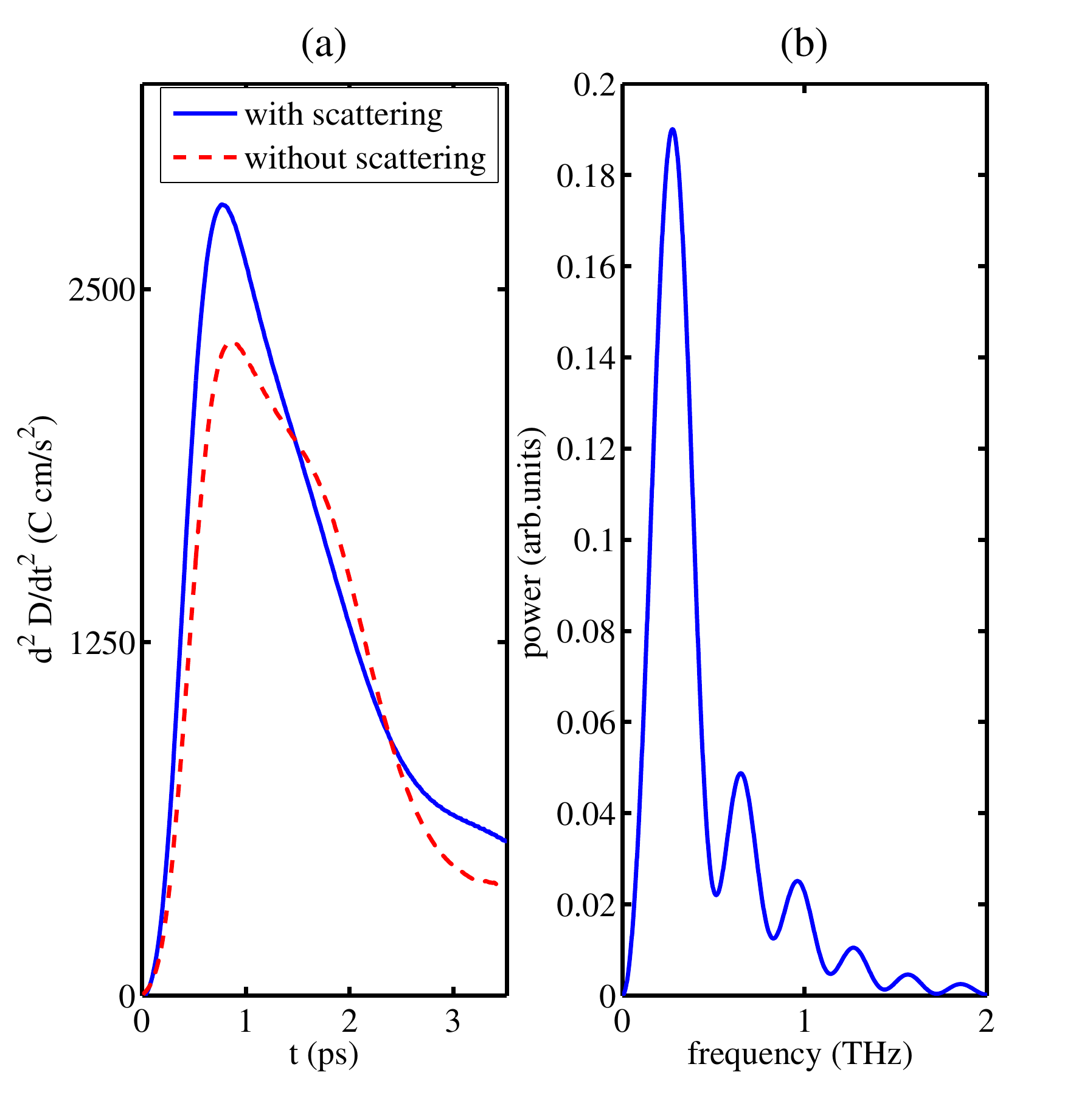}
\end{center}
\caption{(a) Second time derivative of the dipole moment vs time and (b) power spectrum for $N=250$ electrons/holes. 
Scattering times are 2 ps for both types of carriers.} \label{Figure9}
\end{figure}

\section{Appendix} 
Here we derive the expression for electric field and show the importance of the 
points where the derivative of the total charge density vanishes. 
Electric field is derived from the Coulomb forces for a singe-wall nanotube:

$$
E(x,t) = \int_{-\infty}^{\infty}
\left[\bar{p}(x-s,t) -\bar{n}(x-s,t)\right]{\cal K}(s)\,{d}s,
\eqno{\rm (A1)}
\label{E}
$$
with
$$
 {\cal K}(s) =  -\frac{2e}{\pi\epsilon_\perp}
\int_0^{\pi/2}\frac{s\,{d}\theta}{\left(s^2+4 a^2\displaystyle{\frac{\epsilon_\|}{\epsilon_\perp}}\sin^2\theta\right)^{3/2}}.
\eqno{\rm (A2)}
$$
We describe ensemble of nanotubes of the radius $a$ as an anisotropic  medium with 
the longitudinal and transversal permittivities  $\epsilon_\|$ and $\epsilon_\perp$,
respectively  \cite{Lu,Wehling}. The shape of ${\cal K}(s)$ in Eq. (A2) follows from the Poisson equation 
for the electric potential of a point positive charge $-e$ at the origin, 
$$
[\epsilon_\|\partial_x^2+\epsilon_\perp\nabla_\perp^2]\mathcal{V}=
4\pi e\delta(x)\delta(\mathbf{x}_\perp),\quad \mathbf{x}_\perp=(y,z),
\eqno{\rm (A3)}
$$
which can be written as 
$$
\left(\frac{\partial^2}{\partial \tilde{x}^2}+\frac{\partial^2}{\partial\tilde{y}^2}+\frac{\partial^2}{\partial \tilde{z}^2}\right)
\!\mathcal{V}=\frac{4\pi e}{\epsilon_\perp\sqrt{\epsilon_\|}}\delta(\tilde{x})\delta(\tilde{y})\delta(\tilde{z}), 
\eqno{\rm (A4)}
$$
$$
\quad \tilde{x}=\frac{x}{\sqrt{\epsilon_\|}},\, (\tilde{y},\tilde{z})=\frac{1}{\sqrt{\epsilon_\perp}}(y,z).
$$
In terms of the original variables, the solution is
$$
\mathcal{V}=-\frac{e}{\epsilon_\perp\sqrt{x^2+\displaystyle{\frac{\epsilon_\|}{\epsilon_\perp}}\mathbf{x}_\perp^2}}, 
\eqno{\rm (A5)}
$$
$$
\frac{\partial\mathcal{V}}{\partial x}= 
\frac{ex}{\epsilon_\perp}\left[x^2+\frac{\epsilon_\|}
{\epsilon_\perp}\mathbf{x}_\perp^2\right]^{-3/2}.
\eqno{\rm (A6)}
$$
The electric field (\ref{E}) is found straightforwardly 
by convolution of $-\partial\mathcal{V}/\partial x$ with the  
charge density $[\bar{p}(x,t)-\bar{n}(x,t)]\,\delta(|\mathbf{x}_\perp|-a)/2\pi\,a$ and 
changing variables in the resulting integral. After integrating by parts, 
changing variables and using that the electron and hole densities rapidly decrease at large $|x|\gg\Lambda$, (A1) becomes
\begin{widetext}
$$
E(x,t) = \frac{e}{\pi\sqrt{\epsilon_\|\epsilon_\perp}}\,\frac{\partial}{\partial x}\int_{-\infty}^{+\infty}ds
\left[\bar{p}\left(x-2as\sqrt{\frac{\epsilon_\|}{\epsilon_\perp}},t\right)
- \bar{n}\left(x-2as\sqrt{\frac{\epsilon_\|}{\epsilon_\perp}},t\right)\right]
\int_0^{\pi/2}\frac{{d}\theta}{\left(s^2+\sin^2\theta\right)^{1/2}} 
\eqno{\rm (A7)}
\label{E1}
$$
$$
=\frac{e}{\pi\sqrt{\epsilon_\|\epsilon_\perp}}\,\frac{\partial}{\partial x}\int_{-\infty}^{+\infty}
\left[\bar{p}\left(x-2as\sqrt{\frac{\epsilon_\|}{\epsilon_\perp}},t\right) -
\bar{n}\left(x-2as\sqrt{\frac{\epsilon_\|}{\epsilon_\perp}},t\right)\right]K\!\left(\frac{1}{\sqrt{s^2+1}}\right)
\frac{{d}s}{\sqrt{s^2+1}}. 
\eqno{\rm (A8)}
\label{E2}
$$
\end{widetext}

\noindent Here we have written the integral over $\theta$ in terms of the complete elliptic integral of the first kind.
Taking into account that the length scale of the electron and hole density distributions is of the 
order of $\Lambda\gg a$, one can approximate this expression as
$$
E(x,t)\sim \frac{ef}{\sqrt{\epsilon_\|\epsilon_\perp}}\frac{\partial}{\partial x}[\bar{p}(x,t)-\bar{n}(x,t)],
\eqno{\rm (A9)}
\label{E3}
$$
with numerical factor
\begin{widetext}
$$
f=
\pi\ln 2+\frac{\pi^2}{8}+2\int_0^\infty\!\left[K\!\left(\frac{1}{\sqrt{s^2+1}}\right)\!-\frac{\pi}{2}
-\frac{\pi}{8\sqrt{1+s^2}}\right]\! \frac{ds}{\sqrt{1+s^2}}\approx 5.43
\eqno{\rm (A10)}
$$
\end{widetext}
The approximate expression (A9) is a local relationship between the charge density and electric 
field, thereby depending on the geometric mean of the permittivities, $\sqrt{\epsilon_{\|}\epsilon_{\perp}}$, 
to which both contribute equally.

\end{document}